\documentstyle[aps,epsfig]{revtex}
\newcommand{\gsim}{\mathrel{\rlap{\lower4pt\hbox{\hskip0pt$\sim$}}
\raise2pt\hbox{$>$}}}
\newcommand{\lsim}{\mathrel{\rlap{\lower4pt\hbox{\hskip0pt$\sim$}}
\raise2pt\hbox{$<$}}}
\begin{document}
\twocolumn
\preprint{Preprint Number: \parbox[t]{45mm}{ANL-PHY-9380-TH-99\\
}}
 \title{Diquarks: condensation without bound states}
\author{J. C. R. Bloch, C. D. Roberts and
        S. M. Schmidt\vspace*{0.2\baselineskip}}

 \address{Physics Division, Bldg. 203, Argonne National Laboratory,
 Argonne IL 60439-4843}  
\date{
Preprint: ANL-PHY-9380-TH-99; 
Pacs Numbers: 12.38.Mh, 24.85.+p, 12.38.Lg, 11.10.St}
\maketitle
\begin{abstract}
We employ a bispinor gap equation to study superfluidity at nonzero chemical
potential: $\mu\neq 0$, in two- and three-colour QCD.  The two-colour theory,
QC$_2$D, is an excellent exemplar: the order of truncation of the quark-quark
scattering kernel: $K$, has no qualitative impact, which allows a
straightforward elucidation of the effects of $\mu$ when the coupling is
strong.  In rainbow-ladder truncation, diquark bound states appear in the
spectrum of the three-colour theory, a defect that is eliminated by an
improvement of $K$.  The corrected gap equation describes a superfluid phase
that is semi-quantitatively similar to that obtained using the rainbow
truncation.  A model study suggests that the width of the superfluid gap and
the transition point in QC$_2$D provide reliable quantitative estimates of
those quantities in QCD.
\end{abstract}
\section{Introduction}
In the application of Dyson-Schwinger equations\cite{cdragw} (DSEs) extensive
use has been made of models based on the rainbow-ladder truncation, with
contemporary variants\cite{mr97mt99} providing improved links with QCD.  This
truncation is also implicit in the class of model field theories with
four-fermion interactions, such as the Nambu--Jona-Lasinio (NJL)
model~\cite{hugorev} and the Global Colour Model\cite{peterrev}, which have
been used successfully in describing aspects of the strong interaction.  Such
models admit the construction\cite{regdq} of a meson-diquark auxiliary-field
effective-action, which is important in developing an understanding of
nucleons using the relativistic Fadde'ev equation\cite{faddeev}.  In
addition, it is immediately apparent that the action's steepest-descent
equations admit the possibility of diquark condensation; i.e., quark-quark
Cooper pairing, and that was first explored using a simple version of the NJL
model\cite{kahana}.

A nonzero chemical potential: $\mu\neq 0$, promotes Cooper pairing in fermion
systems, and earlier and independent of these developments in QCD
phenomenology, the possibility that it is exhibited in quark matter was
considered\cite{bl84} using the rainbow-ladder truncation of the gap
equation.  A quark-quark Cooper pair is a composite boson with both electric
and colour charge, and hence superfluidity in quark matter entails
superconductivity and colour superconductivity.  However, the last feature
makes it difficult to identify an order parameter that can characterise a
transition to the superfluid phase: the Cooper pair is gauge dependent and an
order parameter is ideally describable by a gauge-invariant operator.

Determining the $(T,\mu)$ phase diagram of QCD is an important goal.  At
$(T,\mu) = 0$ there is a quark-antiquark condensate: $\langle\bar q
q\rangle\neq 0$, but it is undermined by increasing $T$ and $\mu$, and there
is a domain of the $(T,\mu)$-plane for which $\langle\bar q q\rangle=0$.
Increasing $T$ also opposes Cooper pairing.  However, since increasing $\mu$
promotes it, there may be a (low-$T$,large-$\mu$)-subdomain in which quark
matter exists in a superfluid phase.  That domain may not be accessible at
RHIC, which will concentrate on $\mu\simeq 0$ where all studies indicate that
QCD with two light flavours exhibits a chiral symmetry restoring transition:
$\langle\bar q q\rangle\to 0$, at $T\simeq 150\,$MeV.  However, it may be
discernible in the core of dense astrophysical objects\cite{bl84}, which
could undergo a transition to superfluid quark matter as they cool, and in
baryon-density-rich heavy ion collisions at the BNL-AGS and
CERN-SpS\cite{stephanov}.  An exploration of this possibility using numerical
simulations of lattice-QCD is inhibited by the absence of: (i) a
gauge-independent order parameter for the superfluid phase; and (ii) a
satisfactory procedure for the numerical estimation of an integral with a
complex measure, such as the $\mu\neq 0$ QCD partition function.
Consequently all the information we have comes from models.

The rainbow-ladder truncation has the feature and defect that it generates a
quark-quark scattering kernel, $K$, that is purely attractive in the colour
antitriplet channel, $\bar 3_c$.  It therefore not only yields a $\bar 3_c$
scalar diquark condensate but also $\bar 3_c$ diquark bound
states~\cite{cprdq}; i.e., hitherto unobserved coloured quark-quark bound
states with masses (in GeV)\cite{bsesep}:
\begin{equation}
m_{J^P=0^+}^{ud} = 0.74,\;
m_{1^+}^{ud} = 0.95,\;
m_{0^-}^{ud} = 1.5 = m_{1^-}^{ud}\,.
\end{equation}
($us=ds$ diquarks are also bound; e.g., $m_{0^+}^{us}=0.88$.  Colour-sextet
bound states do not exist because $K$ is purely repulsive in this channel,
even in rainbow-ladder truncation~\cite{cprdq}.)  All models employed to date
in the analysis of quark matter superfluidity have this defect\cite{wilczek},
and we are primarily concerned with the question of whether any model or
truncation with such a flaw can be a reliable tool for exploring
superfluidity in quark matter.  In addressing this issue, it is important to
compare QC$_2$D with QCD because the same mechanism that provides for the
absence of diquark bound states in the latter {\em must} guarantee their
existence in QC$_2$D, where the diquark is the baryon of the theory.  In
fact, it must ensure that flavour-nonsinglet $J^{P=\,\mp}$ mesons are
degenerate with $J^{\pm}$ diquarks\cite{cdrqcII}.

In Sect.~II we describe a bispinor DSE ({\it gap equation}) that is
particularly useful for studying quark and diquark condensation and, in
Sect.~III, employ it in the general analysis of QC$_2$D and also to obtain
quantitative results from a pedagogical model.  In Sect.~IV we focus on QCD,
and employ the model's analogue to exemplify the gap equation and its
solution in rainbow truncation, and also when a $1/N_c$-suppressed
dressed-ladder vertex correction is included.  We summarise and conclude in
Sect.~V.

\section{A Gap Equation}
A direct means of determining whether a SU$_c(N$) gauge theory supports $0^+$
diquark condensation is to study the gap equation satisfied by\footnote{In
our Euclidean formulation: $\{\gamma_\mu,\gamma_\nu\}=2\delta_{\mu\nu}$,
$\gamma_\mu^\dagger = \gamma_\mu$, $p\cdot q=\sum_{i=1}^4 p_i q_i$, and
tr$_D[\gamma_5\gamma_\mu\gamma_\nu\gamma_\rho\gamma_\sigma]=
-4\,\epsilon_{\mu\nu\rho\sigma}$, $\epsilon_{1234}= 1$.}
\begin{eqnarray}
\label{sinv}
\lefteqn{{\cal D}(p,\mu) := }\\
&& \nonumber
{\cal S}(p,\mu)^{-1} =\left(
\begin{array}{cc}
D(p,\mu) & \Delta^i(p,\mu)\, \gamma_5  \lambda_{\wedge}^i \\
 -\Delta^i(p,-\mu)\, \gamma_5  \lambda_{\wedge}^i
        & C D(-p,\mu)^{\rm T} C^\dagger
\end{array}\right),
\end{eqnarray}
where, with $\omega_{[\mu]}= p_4+i\mu$,
\begin{eqnarray}
\label{DCA}
\lefteqn{D(p,\mu)  = }\\
\nonumber && i \vec{\gamma}\cdot \vec{p}\,
A(\vec{p}\,^2,\omega_{[\mu]}^2) + 
i \gamma_4 \,\omega_{[\mu]} \,C(\vec{p}\,^2,\omega_{[\mu]}^2 ) + 
B(\vec{p}\,^2,\omega_{[\mu]}^2 ),
\end{eqnarray}
$\{\lambda_{\wedge}^i$, $i=1\ldots n^\wedge_c$, $n^\wedge_c= N_c (N_c-1)/2\}$
are the antisymmetric generators of SU$_c(N_c)$, and $C= \gamma_2\gamma_4$ is
the charge conjugation matrix: 
$
C\gamma_\mu^{\rm T}C^\dagger = -\gamma_\mu\,;\;
[C,\gamma_5]=0\,.
$
Using the gap equation to study superfluidity makes unnecessary a truncated
bosonisation, which in all but the simplest models is a procedure difficult
to improve systematically.

In addition to the usual colour, Dirac and isospin indices carried by the
elements of ${\cal D}(p,\mu)$, the explicit matrix structure in
Eq.~(\ref{sinv}) exhibits the quark bispinor index and is made with reference
to
\begin{eqnarray}
\label{QQCD}
Q(x) & := & \left(\begin{array}{c}
                        q(x)\\
                        \underline q(x):= \tau^2_f\, C\, \bar q^{\rm T}
                         \end{array} \right)\,,\\
\label{bQQCD}
\bar Q(x) & := &\left(\begin{array}{cc}
                        \bar q(x)\;\;
                        \bar {\underline q}(x):= q^{\rm T} \,C\,\tau^2_f
\end{array} \right),
\end{eqnarray}
where $\{\tau_f^i: i=1,2,3\}$ are Pauli matrices that act on the isospin
index.  Herein we only consider two-flavour theories, SU$_f(N_f=2$), because
$N_f$ does not affect the question at the core of our study, and focus on
$T=0$, since nonzero $T$ can only act to eliminate a condensate.  A nonzero
quark condensate: $\langle \bar q q \rangle \neq 0$, is represented in the
solution of the gap equation by $B(\vec{p}\,^2,\omega_{[\mu]}^2 )\not\equiv
0$ while diquark condensation is characterised by $\Delta^i(p,\mu)\not\equiv
0$, for at least one $i$.

The bispinor DSE can be written in the form
\begin{eqnarray}
\label{bispinDSE}
\lefteqn{{\cal D}(p,\mu)  = }\\
&& \nonumber  {\cal D}_0(p,\mu) 
+ \left( 
\begin{array}{cc}
\Sigma_{11}(p,\mu) & \Sigma_{12}(p,\mu)\\
\gamma_4\,\Sigma_{12}(-p,\mu)\,\gamma_4 & C \Sigma_{11}(-p,\mu)^{\rm T}
C^\dagger  
\end{array} \right),
\end{eqnarray}
where in the absence of a diquark source term
\begin{eqnarray}
{\cal D}_0(p,\mu) 
%
& = & (i\gamma\cdot p + m )\tau_Q^0 - \mu\,\tau_Q^3\,,
\end{eqnarray}
with $m$ the current-quark mass.  Here we have introduced additional Pauli
matrices: $\{\tau_Q^\alpha,\alpha = 0,1,2,3\}$ with $\tau^0 = {\rm
diag}(1,1)$, that act on the bispinor indices.  The structure of
$\Sigma_{ij}(p,\mu)$ specifies the theory and, in practice, also the
approximation or truncation of it.

\section{Two Colours}
\label{SU2}
As an important and instructive first example we consider QC$_2$D.  In this
special case
$ \Delta^i \lambda_\wedge^i =\Delta \tau^2_c $
in Eq.~(\ref{sinv}) and it is useful to employ a modified bispinor
\begin{eqnarray}
Q_2(x) &:= &\left(\begin{array}{c}
                         q(x)\\
\underline q_2:=\tau^2_c\,\underline q(x)  \end{array} \right),\;
%
\end{eqnarray}
with $\bar Q_2$ the obvious analogue of Eq.~(\ref{bQQCD}), so that the
Lagrangian's fermion-gauge-boson interaction term is simply
$ \bar Q_2(x) \,\case{i}{2} g \gamma_\mu \tau_c^k\tau_Q^0
\,Q_2(x)\,A_\mu^k(x) $
because SU$_c(2)$ is pseudoreal; i.e.,  
$ \tau^2_c\left(-\vec{\tau}_c\right)^{\rm T}\tau^2_c = \vec{\tau}_c \,,\;$ 
and the fundamental and conjugate representations are equivalent.

The gap equation at arbitrary order in the systematic, Ward-Takahashi
identity preserving truncation scheme of Ref.\cite{brs96} is readily derived.
For $\mu=0$: $C=A$ in Eq.~(\ref{DCA}), all the functions in the
dressed-bispinor propagator are real and the rainbow truncation yields the
gap equation
\begin{eqnarray}
\label{srainbow}
\lefteqn{{\cal D}(p)= }\\
&& \nonumber i\gamma\cdot p + m 
+ \int\case{d^4 q}{(2\pi)^4} \, g^2 D_{\mu\nu}(p-q)\,
\gamma_\mu \frac{\tau_c^k}{2}\, {\cal S}(q) \,
\gamma_\nu \frac{\tau_c^k}{2}\,.
\end{eqnarray}
Renormalisation is straightforward, even at $\mu\neq 0$\cite{greg}, however,
since it is not relevant to our central theme, we neglect it here.  To solve
Eq.~(\ref{srainbow}) we consider a generalisation\cite{reg85} of
Eq.~(\ref{sinv})
\begin{eqnarray}
\label{sinvgen2}
{\cal D}(p) & = &
i\gamma\cdot p\, A(p^2) + 
{\cal V}(-\mbox{\boldmath$\pi$})\, {\cal M}(p^2)\,;\\
{\cal V}(\mbox{\boldmath$\pi$}) & = &
\exp\left\{i \gamma_5\, \sum_{\ell=1}^5\,T^\ell\, \pi^\ell(p^2)\right\}
= {\cal V}(-\mbox{\boldmath$\pi$})^{-1} \,,
\end{eqnarray}
where $\{T^{1,2,3}= \tau_Q^3 \otimes \vec{\tau_f},\, T^4= \tau_Q^1\otimes
\tau_f^0,\, T^5= \tau_Q^2\otimes\tau_f^0 \}$, $\{T^i,T^j\}=2 \delta^{ij}$, so
that
\begin{eqnarray} 
\label{Sp}
{\cal S}(p) & =&  \frac{-i\gamma\cdot p A(p^2) + 
{\cal V}(\mbox{\boldmath$\pi$})  {\cal M}(p^2)} 
{p^2 A^2(p^2) + {\cal M}^2(p^2)}\,,\\
& := & -i\gamma\cdot p \,\sigma_V(p^2) + 
{\cal V}(\mbox{\boldmath$\pi$})\,\sigma_S(p^2)\, . 
\end{eqnarray}
($\mbox{\boldmath$\pi$}=(0,0,0,0,-\case{1}{4}\pi)$ produces an inverse
bispinor propagator with the simple form in Eq.~(\ref{sinv}).)

For $\mbox{\boldmath$\pi$}(p^2)=\,$constant, substituting
Eq.~(\ref{sinvgen2}) into Eq.~(\ref{srainbow}), yields
\begin{eqnarray}
\lefteqn{\gamma\cdot p \,[A(p^2) -1 ]= }\\
&& \nonumber - \int\case{d^4 q}{(2\pi)^4} \, 
g^2 D_{\mu\nu}(p-q)\,
\gamma_\mu \frac{\tau_c^k}{2}\, 
\gamma\cdot q\, \sigma_V(q^2)\,
\gamma_\nu \frac{\tau_c^k}{2}\, , \\
\label{eqM}
\lefteqn{{\cal M}(p^2)  - m\,{\cal V}(\mbox{\boldmath$\pi$}) =    }\\
&& \nonumber
\int\case{d^4 q}{(2\pi)^4} \, g^2 D_{\mu\nu}(p-q)\,
\gamma_\mu \frac{\tau_c^k}{2}\, 
\sigma_S(q^2)\, 
\gamma_\nu \frac{\tau_c^k}{2}\,  .
\end{eqnarray}
It is clear from these equations that the gap equation in rainbow truncation
is {\em independent} of \mbox{\boldmath$\pi$} in the chiral limit.  As this
result is true order-by-order in the truncation scheme of Ref.~\cite{brs96},
it is a general property of the complete QC$_2$D gap equation.  Hence, if the
interaction is strong enough to generate a mass gap, then that gap describes
a five-parameter continuum of degenerate condensates:
\begin{equation}
\label{cndst}
\langle \bar Q_2 
{\cal V}(\mbox{\boldmath$\pi$}) Q_2\rangle \neq 0\,,
\end{equation}
and there are 5 associated Goldstone bosons: 3 pions, a diquark and an
anti-diquark, which is a well-known consequence of the Pauli-G\"ursey symmetry
of QC$_2$D.

For $m\neq 0$, it is clear from Eq.~(\ref{eqM}) that the gap equation
requires
${\rm tr}_{fQ} \left[ T^i {\cal V} \right] = 0\,,$
i.e., in this case only $\langle \bar Q_2 Q_2 \rangle \neq 0$.  The Goldstone
bosons are now massive but remain degenerate.

The Landau gauge dressed-gauge-boson propagator is
\begin{equation}
g^2 D_{\mu\nu}(k) = \left(\delta_{\mu\nu}-\frac{k_\mu k_\nu}{k^2}\right) 
                {\cal F}(k^2)
\end{equation}
and to explore $\mu\neq 0$ we employ a pedagogical model for the vacuum
polarisation in QC$_2$D:
\begin{equation}
\label{delta}
{\cal F}_2(k^2)= \case{64}{9}\pi^4\, {\hat \eta}^2\,\delta^4(k)\,.
\end{equation}
This form was introduced\cite{mn83} for the modelling of confinement in QCD.
However, it is also appropriate here because the string tension in QC$_2$D is
nonzero, and that is represented implicitly in Eq.~(\ref{delta}) via the
mass-scale ${\hat \eta}$.  Further, a simple extension of the model has been
used efficaciously\cite{thermo,basti} as a heuristic tool for the analysis of
QCD at nonzero-$(T,\mu)$.  The mass of the model's $J=1$ composites is a
useful reference scale and for $m=0$ in rainbow-ladder truncation
\begin{equation}
m_{J=1}^2 = \case{1}{2}\,{\hat \eta}^2\,.
\end{equation}
$m_{J=1}$ is weakly dependent on $m$, changing by $\lsim 2$\% on
$m\in[0,0.01]\,{\hat \eta}$, while adding $1/N_c$-suppressed vertex
corrections produces an increase of $<10$\%\cite{brs96}.

We now consider the $\mu\neq 0$ gap equation and suppose a solution of the
form
\begin{eqnarray}
\label{dqD}
{\cal D}(p,\mu) & = & \left(
\begin{array}{cc}
D(p,\mu) & \gamma_5\,\Delta(p,\mu)  \\
- \gamma_5\Delta^\ast(p,\mu) & \tilde D(p,\mu)
\end{array}
\right)\,,
%
\end{eqnarray}
with $D(p,\mu)$ defined in Eq.~(\ref{DCA}) and $\tilde D(p,\mu):= C
\,D(-p,\mu)\,C^\dagger$.  In the absence of a diquark condensate; i.e., for
$\Delta \equiv 0$,
\begin{equation}
[U_B(\alpha), {\cal D}(p,\mu)]=0\,, \; 
U_B(\alpha):= {\rm e}^{i \alpha \tau_Q^3 \otimes \tau_f^0}\,,
\end{equation}
which is a manifestation of baryon number conservation in QC$_2$D.  

The inverse, ${\cal S}(p,\mu)$, is sufficiently complicated that it provides
little insight directly.  However, that can be obtained using
Eq.~(\ref{delta}), which yields an algebraic gap equation.  Using the rainbow
truncation we find, at $p^2=|\vec{p}|^2 + p_4^2 =0$:
\begin{eqnarray}
A - 1 & = & \case{1}{2}\,{\hat \eta}^2 \,K\,\left\{
                A \,(B^{\ast 2} - C^{\ast 2} \mu^2)
                + A^\ast \,|\Delta|^2 \right\}\,,\\
\mu(C - 1) & = & \case{\mu}{2}\,{\hat \eta}^2 \,K\,\left\{ C\,(B^{\ast 2} -
                C^{\ast 2} \mu^2) -C^\ast\,|\Delta|^2 \right\}\,,\\
B - m & = & {\hat \eta}^2 \,K\,\left\{
                B\,(B^{\ast 2} - C^{\ast 2} \mu^2)
                + B^\ast \,|\Delta|^2 \right\}\,,\\
\Delta & = & {\hat \eta}^2 \,K\, \left\{
                \Delta\,(|B|^2 + |C|^2 \mu^2)
                +\Delta \,|\Delta|^2 \right\}\,,
\end{eqnarray}
with 
$K^{-1}  =  |B^2-C^2\mu^2|^2 + 2 |\Delta|^2 (|B|^2+ |C|^2 \mu^2) +
|\Delta|^4\,. $
These equations illustrate: the $B \leftrightarrow \Delta$ degeneracy
described above for $(m,\mu)=0$; that $\Delta$ is real for all $\mu$; and
also the action of $\mu$: enhancing the coupling in the $\Delta$ equation but
suppressing it for $B$, which is how increasing $\mu$ promotes diquark
condensation at the expense of the quark condensate.

For $(m,\mu)=0$ the rainbow gap equation is Eq.~(\ref{srainbow}) and with
Eq.~(\ref{delta}) the solution is:
\begin{eqnarray}
\lefteqn{{\cal M}^2(p^2)  :=  }\\
&& \nonumber B^2(p^2) + \Delta^2(p^2) =  \left\{\begin{array}{ll}
{\hat \eta}^2 - 4 p^2, & p^2 < \frac{{\hat \eta}^2}{4}\\
0,              & {\rm otherwise}\;,
\end{array}\right.\\
\label{ACeqn}
\lefteqn{A(p^2)  =  C(p^2)   }\\
&& \nonumber 
= \left\{\begin{array}{ll}
2, & p^2 < \frac{{\hat \eta}^2}{4}\\
\case{1}{2}\left( 1 + \sqrt{1 + \frac{2{\hat \eta}^2}{p^2}}\right), 
   & {\rm otherwise}\;.
\end{array}\right.
\end{eqnarray}
The dynamically generated mass function, ${\cal M}(p^2)$, is tied to the
existence of quark and/or diquark condensates and breaks chiral symmetry.
Further, in {\em combination} with the momentum-dependent vector self energy,
it ensures that the quark propagator does not have a Lehmann representation
and hence can be interpreted as describing a confined
quark~\cite{cdragw,gastao}.  The interplay between the scalar and vector self
energies is the key to realising confinement in this way~\cite{brw92}, and is
precluded in studies that discard the vector self energy.

For $\mu\neq 0$ and arbitrary $p$ we solve the rainbow gap equation
numerically, and determine whether quark or diquark condensation is stable by
evaluating
\begin{equation}
\label{deltaP}
\delta P := P(\mu,{\cal S}[B=0,\Delta]) - P(\mu,{\cal S}[B,\Delta=0])\,,
\end{equation} 
where the pressure is calculated using a steepest descent
approximation\cite{haymaker}:
\begin{equation}
P[{\cal S}]= - {\rm TrLn[{\cal S}]}
        - \case{1}{2}{\rm Tr}[ ({\cal D}-{\cal D}_0) {\cal S}]\,.
\end{equation}
$\delta P>0$ indicates that diquark condensation is favoured.  

The calculation of $\delta P$ is facilitated by employing the $\mu$-dependent
``bag constants''\cite{reg85}
\begin{eqnarray}
\lefteqn{{\cal B}_B(\mu) := }\\
&& \nonumber P(\mu,{\cal S}[B,\Delta=0]) - P(\mu,{\cal S}[B=0,\Delta=0])\,,
%
\end{eqnarray}
with ${\cal B}_\Delta(\mu)$ an obvious analogue.  They measure the stability
of a quark- or diquark-condensed vacuum relative to that with chiral symmetry
realised in the Wigner-Weyl mode.  The $(m,\mu)=0$ degeneracy of the quark
and diquark condensates is manifest in
\begin{equation}
{\cal B}_B(0) = {\cal B}_\Delta(0) = (0.092\,{\hat \eta})^4 
= (0.13 \,m_{J=1})^4\,.
\end{equation}

Increasing $\mu$ at $m=0$ and excluding diquark condensation one
finds\cite{thermo} chiral symmetry restoration at
\begin{equation}
\mu_{2c}^{B,\Delta=0} = 0.28\, {\hat \eta}
\end{equation}
when ${\cal B}_B(\mu) = 0$; i.e., the pressure of the Wigner and
quark-condensed phases are equal.  However,
${\rm for~all} \;\mu > 0$: $\delta P>0$, ${\cal B}_\Delta(\mu)>0$,
with 
${\cal B}_\Delta(\mu_{2c}^{B,\Delta=0}) = (0.20\, m_{J=1})^4 > {\cal
B}_\Delta(0)$. 
Therefore the vacuum is unstable with respect to diquark condensation for all
$\mu> 0$ and one has confinement and dynamical chiral symmetry breaking to
arbitrarily large values.  Of course, we have ignored the possibility that
${\hat \eta}$ is $\mu$-dependent.  In a more realistic model, the
$\mu$-dependence of ${\hat \eta}$ would be significant in the vicinity of
$\mu_{2c}^{B,\Delta=0}$, with ${\hat \eta}\to 0$ as $\mu\to \infty$, which would
ensure deconfinement and chiral symmetry restoration at large-$\mu$.

$\Delta\neq 0$ in Eq.~(\ref{dqD}) corresponds to
$\mbox{\boldmath$\pi$}=(0,0,0,0,\case{1}{2}\pi)$ in Eq.~(\ref{cndst}); i.e.,
$ \langle \bar Q_2 i\gamma_5\tau_Q^2 Q_2\rangle \neq 0\,, $
and although the $\mu\neq 0$ theory is invariant under
$ Q_2 \to U_B(\alpha) \,Q_2\,,\; \bar Q_2 \to \bar Q_2\,U_B(-\alpha) \,, $
which is associated with baryon number conservation, the diquark condensate
breaks this symmetry:
\begin{eqnarray}
\lefteqn{\langle \bar Q_2 i\gamma_5\tau_Q^2 Q_2\rangle}\\
&& \nonumber \to 
\cos(2 \alpha)\, \langle \bar Q_2 i\gamma_5\tau_Q^2 Q_2\rangle
- \sin(2 \alpha)\,\langle \bar Q_2 i\gamma_5\tau_Q^1 Q_2\rangle\,.
\end{eqnarray}
Hence, for $(m=0,\mu\neq 0)$, one Goldstone mode remains.

For $m\neq 0$ and small values of $\mu$, the gap equation only admits a
solution with $\Delta \equiv 0$; i.e., diquark condensation is blocked.
However, with increasing $\mu$ a diquark condensate is generated; e.g., we
find the following minimum chemical potentials for diquark condensation
\begin{equation}
\begin{array}{lcl}
m= 0.013\,m_{J=1} & \Rightarrow & \mu^{\Delta \neq 0} =
0.051\,m_{J=1}\,,\\
m= 0.13\,m_{J=1} & \Rightarrow & \mu^{\Delta \neq 0} = 0.092\,m_{J=1}\,.
\end{array}
\end{equation}

Improving on rainbow-ladder truncation may yield quantitative changes of
$\lsim 20$\% in the illustrative results provided by our model of QC$_2$D.
However, the pseudoreality of QC$_2$D and the equal dimension of the colour
and bispinor spaces, which underly the theory's Pauli-G\"ursey symmetry,
ensure that the entire discussion remains qualitatively unchanged.  QCD,
however, has two significant differences: the dimension of the colour space
is greater than that of the bispinor space and the fundamental and conjugate
representations of the gauge group are not equivalent.  The latter is of
obvious importance because it entails that the quark-quark and
quark-antiquark scattering matrices are qualitatively different.

\section{Three Colours}
In canvassing superfluidity in QCD we choose $\Delta^i\lambda^i_\wedge =
\Delta\, \lambda^2$ in Eq.~(\ref{sinv}) so that 
\begin{eqnarray}
\label{sinvQCD}
\lefteqn{{\cal D}(p,\mu) = }\\
&& \nonumber \left(
\begin{array}{c|c}
D_\|(p,\mu) P_\| + D_\perp(p,\mu) P_\perp 
        & \Delta(p,\mu) \gamma_5 \lambda^2 \\ \hline
- \Delta(p,-\mu) \gamma_5 \lambda^2 \mbox{\rule{0mm}{1.2\parindent}}
        & \tilde D_\|(p,\mu) P_\| + \tilde D_\perp(p,\mu) P_\perp 
\end{array}\right)
\end{eqnarray}
where $P_\|=(\lambda^2)^2$, $P_\perp + P_\| = {\rm diag}({1,1,1})$, and $D_\|
\ldots \tilde D_\perp$ are defined via obvious generalisations of
Eqs.~(\ref{sinv}) and (\ref{DCA}).  The evident, demarcated block structure
makes explicit the bispinor index.  Here each block is a $3\times 3$ colour
matrix and the subscripts: $\|$, $\perp$, indicate whether or not the
subspace is accessible via $\lambda_2$.  The bispinors associated with this
representation are given in Eqs.~(\ref{QQCD}) and (\ref{bQQCD}), and in this
case the Lagrangian's quark-gluon interaction term is
$ \bar Q(x) i g\Gamma_\mu^a Q(x) A_\mu^a(x)$, 
\begin{eqnarray}
\label{bispingamma}
\Gamma_\mu^a & = & \left(\begin{array}{c|c}
        \case{1}{2}\gamma_\mu \lambda^a 
                & 0 \\[0.2\parindent]\hline
        0 \mbox{\rule{0mm}{0.8\parindent}}
                & -\case{1}{2}\gamma_\mu (\lambda^a)^{\rm T}\end{array}\right)\,.
\end{eqnarray}
It is again straightforward to derive the gap equation at arbitrary order in
the truncation scheme of Ref.\cite{brs96} and it is important to note that
because
\begin{eqnarray}
\lefteqn{D_\|(p,\mu) P_\| + D_\perp(p,\mu) P_\perp 
= \lambda^0 \left\{\case{2}{3} D_\|(p,\mu) \right.}\\
&& \nonumber
\left. + \case{1}{3} D_\perp(p,\mu)
\right\}
+ \case{1}{\sqrt{3}}\lambda^8 
\left\{ D_\|(p,\mu) -  D_\perp(p,\mu)\right\}
\end{eqnarray}
the interaction: $\Gamma_\mu^a {\cal S}(p,\mu)\Gamma_\nu^a$, necessarily
couples the $\|$- and $\perp$-components.  That interplay is discarded in
models that ignore the vector self energy of quarks, which is a necessary and
qualitatively important feature of QCD~\cite{thermo,basti,brw92,kisslinger}.

\subsection{Rainbow truncation}
Diquark condensation at $\mu=0$ was studied in Ref.\cite{trento} using a
minor quantitative adjustment of the confining model gluon propagator defined
via Eq.~(\ref{delta}):
\begin{equation}
\label{delta3}
{\cal F}(k^2)= 4\pi^4\, \eta^2\,\delta^4(k)\,,
\end{equation}
with which the rainbow-truncation gap equation is
\begin{eqnarray}
\label{dseRT}
{\cal D}(p,\mu) & = & {\cal D}_0(p,\mu) 
+ \case{3}{16}\eta^2\,\Gamma_\rho^a \,{\cal S}(p,\mu)\, \Gamma_\rho^a\,.
\end{eqnarray}
Solving this and the ladder-truncation Bethe-Salpeter equation one
obtains\cite{thermo,basti}
\begin{eqnarray}
\label{resRL1}
m_\omega^2 = m_\rho^2 & = & \case{1}{2}\,\eta^2\,,\\
\langle \bar q q \rangle & = & (0.11\,\eta)^3\,,\\
\label{resRL3}
{\cal B}_B(\mu=0) & = & (0.10\,\eta)^4\,,
\end{eqnarray}
and momentum-dependent vector self energies, Eq.~(\ref{ACeqn}), which lead to
an interaction between the $\|$- and $\perp$-components of ${\cal D}$ that
blocks diquark condensation.  This 
\begin{figure}[t]
\centering{\ \epsfig{figure=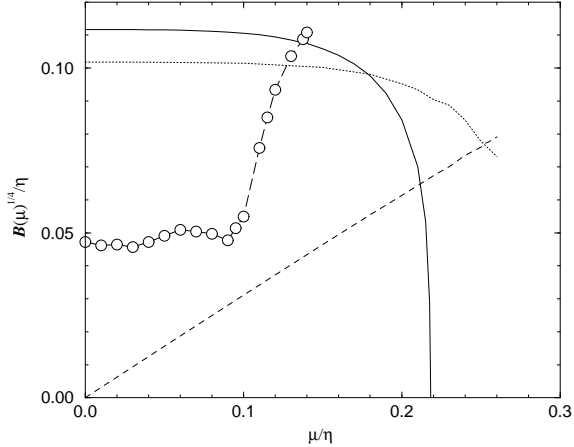,height=6.0cm}}
\caption{$\mu$-dependent ``bag-constants'' in the QCD model defined via
Eq.~(\protect\ref{delta3}).  Rainbow truncated gap equation -- Dotted line:
${\cal B}_B(\mu)$, short-dashed line: ${\cal B}_\Delta(\mu)$.  At the
intersection, where the system flips to the superfluid phase, ${\cal
B}_\Delta(\mu_c^{B=0,\Delta}) = (0.75)^4\,{\cal B}_B(0)$.  Vertex-corrected
gap equation -- Solid line: ${\cal B}_B(\mu)$, long-dashed line with circles:
${\cal B}_\Delta(\mu)$.  At the intersection: ${\cal B}_\Delta =
(0.96)^4\,{\cal B}_B(0)$.  The structure evident in ${\cal B}_\Delta(\mu)$ is
an artefact characteristic of
Eq.~(\protect\ref{delta3})\protect\cite{thermo}.
\label{bagconstants}}
\end{figure}
\hspace*{-\parindent}is in spite of the fact that
$ \lambda^a \lambda^2 (-\lambda^a)^{\rm T} = \case{1}{2}\lambda^a
\lambda^a\,, $
which entails that the ladder-truncation quark-quark scattering kernel is
purely attractive and strong enough to produce diquark bound
states~\cite{cprdq}.

For $\mu\neq 0$ and in the absence of diquark condensation, the model defined
via Eq.~(\ref{delta3}) exhibits\cite{thermo} coincident, first order chiral
symmetry restoring and deconfining transitions at
\begin{equation}
\mu_{c,\,{\rm rainbow}}^{B,\Delta=0} = 0.28 \,\eta\,,
\end{equation}
which is where ${\cal B}_B=0$.  However, for $\mu\neq 0$ Eq.~(\ref{dseRT})
admits a solution with $\Delta(p,\mu)\not\equiv 0$ and $B(p,\mu)\equiv 0$.
$\delta P$ in Eq.~(\ref{deltaP}) again determines whether the quark-condensed
or superfluid phase is the stable ground state.  With increasing $\mu$,
${\cal B}_B(\mu)$ decreases, very slowly at first, and ${\cal
B}_{\Delta}(\mu)$ increases rapidly from zero.  As illustrated in
Fig.~\ref{bagconstants}, that evolution continues until
\begin{equation}
\mu_{c,\,{\rm rainbow}}^{B=0,\Delta} = 0.25 \,\eta = 0.89\,\mu_{c,\,{\rm
rainbow}}^{B,\Delta=0}\,,
\end{equation}
where ${\cal B}_\Delta(\mu)$ becomes greater-than ${\cal B}_B(\mu)$.  This
signals a first order transition to the superfluid ground state\footnote{
With $\eta$ independent of $\mu$, quark confinement, expressed as the absence
of a Lehmann representation for ${\cal S}$, persists in the superfluid
domain.}
and at the boundary
\begin{equation}
\label{qqqbq}
\langle \bar Q i\gamma_5\tau_Q^2\lambda^2 Q\rangle_{\mu=\mu_{c,\,{\rm
rainbow}}^{B=0,\Delta}} 
= (0.65)^3\,\langle \bar Q Q\rangle_{\mu=0}\,.
\end{equation}
These results are typical\cite{biel} of rainbow truncation models in which
the parameters in the dressed-gluon propagator are tuned to yield the correct
$\pi$-$\rho$ mass splitting.  
\begin{figure}[t]
\centering{\ \epsfig{figure=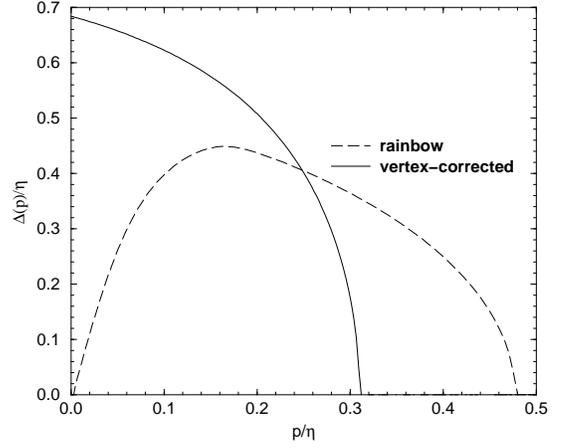,height=6.0cm}}
\caption{Dashed line: $\Delta(z,\mu_c^{B,\Delta})$ obtained in rainbow
truncation with the QCD model defined via Eq.~(\protect\ref{delta3}), plotted
for $\alpha = 0$ as a function of $p$, where $z= p\,(0,0,\sin\alpha,i\mu+
\cos\alpha )$.  As $\mu$ increases, the peak position shifts to larger values
of $p$ and the peak height increases.  Solid line: $\Delta(z,\mu=0)$ obtained
as the solution of Eq.~(\protect\ref{dseNO}), the vertex-corrected gap
equation, also with $\alpha=0$.
\label{deltak}}
\end{figure}
\hspace*{-\parindent}The solution of the rainbow gap equation:
$\Delta(p,\mu_c^{B,\Delta})$, which is real and characterises the diquark
gap, is plotted in Fig.~\ref{deltak}.  It vanishes at $p^2=0$ as a
consequence of the $\|$-$\perp$ coupling that blocked diquark condensation at
$\mu=0$, and also at large $p^2$, which is the manifestation of asymptotic
freedom in the model.

\subsection{Vertex corrected gap equation}
\begin{figure}[t]
\centering{\ \epsfig{figure=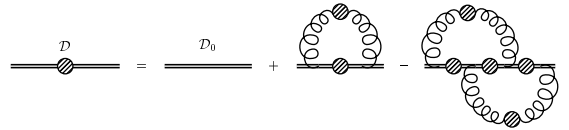,height=1.70cm}}\vspace*{0.5\baselineskip}

\caption{Illustration of the dressed-ladder vertex-corrected gap equation,
Eq.~(\protect\ref{dseNO}).  Each bispinor quark-gluon vertex is bare, given
by Eq.~(\protect\ref{bispingamma}).
\label{dressed}}
\end{figure}

The next-order term in the gap equation corresponds to adding a
$1/N_c$-suppressed dressed-ladder correction to the quark-gluon vertex, and
using Eq.~(\ref{delta3}) this yields
\begin{eqnarray}
\label{dseNO}
\lefteqn{{\cal D}(p,\mu)  =  {\cal D}_0(p,\mu) 
+ \case{3}{16}\eta^2\,\Gamma_\rho^a \,{\cal S}(p,\mu)\, \Gamma_\rho^a}\\
&& \nonumber 
- \case{9}{256}\eta^4\,\Gamma_\rho^a\, {\cal S}(p,\mu)\,
         \Gamma_\sigma^b\, {\cal S}(p,\mu)\,
                \Gamma_\rho^a\, {\cal S}(p,\mu)\, \Gamma_\sigma^b\,,
\end{eqnarray}
which is illustrated in Fig.~\ref{dressed}.  
The kernel of the Bethe-Salpeter equation receives three additional
contributions at this order.  Their net effect is repulsive at timelike total
momentum and hence they prevent the formation of diquark bound
states\cite{brs96,Hellstern}.  The $\eta^4$ term in Eq.~(\ref{dseNO}) means
that an algebraic solution cannot be obtained, however, a numerical solution
is possible.  For simplicity we only consider $m=0$ since $m\neq 0$ opposes
diquark condensation, as we saw in Sect.~\ref{SU2}.  At this order there is a
$\Delta\not \equiv 0$ solution even for $\mu=0$, which is illustrated in 
\hspace*{-\parindent}Fig.~\ref{deltak}, and
\begin{eqnarray}
m_\rho^2 & = & (1.1)^2 \,m_\rho^{2\;{\rm ladder}}\\
\langle \bar Q Q\rangle & = & (1.0)^3\,\langle \bar Q Q\rangle^{\rm
rainbow}\\
{\cal B}_B & = & (1.1)^4\, {\cal B}_B^{\,{\rm rainbow}}\,,
\end{eqnarray}
where the rainbow-ladder results are given in
Eqs.~(\ref{resRL1})-(\ref{resRL3}), and
\begin{eqnarray}
\langle \bar Q i\gamma_5\tau_Q^2\lambda^2 Q\rangle 
& = & (0.48)^3\,\langle \bar Q Q\rangle\,,\\
{\cal B}_\Delta& = & (0.42)^4\,{\cal B}_B \,.
\end{eqnarray}
Unsurprisingly the quark-condensed phase is favoured.  Precluding diquark
condensation, the model exhibits coincident, first order chiral symmetry
restoring and deconfinement transitions\footnote{
At $\mu=0$ and $T\neq 0$ these transitions are second order, and the critical
temperature is reduced by $<2$\% when calculated using Eq.~(\ref{dseNO})
instead of the rainbow truncation\cite{arne}.  }
at
\begin{equation}
\mu_c^{B,\Delta=0} = 0.77\,\mu_{c\,{\rm rainbow}}^{B,\Delta=0}\,.
\end{equation}
 
Our numerical results\footnote{
Following the evolution of the diquark gap with increasing $\mu$ is
numerically difficult because of the interaction between the $\|$- and
$\perp$-components of ${\cal D}$.  With that interaction the gap equation
yields nine coupled quintic algebraic equations in nine variables, and of the
many possible solutions one must follow the correct branch as $p$ and $\mu$
evolve.}
for the $\mu$-dependence of the ``bag constants'' are depicted in
Fig.~\ref{bagconstants}, which shows there is a transition to the superfluid
phase at
\begin{equation}
\mu_{c}^{B=0,\Delta} =  0.63\, \mu_{c}^{B,\Delta=0}\,,
\end{equation}
and at the boundary (cf. Eq.~(\ref{qqqbq}))
\begin{equation}
\langle \bar Q i\gamma_5\tau_Q^2\lambda^2 Q
\rangle_{\mu = 0.63\,\mu_{c}^{B,\Delta=0}}
= (0.51)^3\,\langle \bar Q Q\rangle_{\mu=0}\,.
\end{equation}
The ratio of the condensates increases by $<7$\% on
$\mu\in[0,\mu_{c}^{B=0,\Delta}]$.  Quantitatively, the next-order correction
leads to a reduction in the critical chemical potential for the transition to
superfluid quark matter but doesn't much affect the width of the gap.
Qualitatively, the transition occurs despite there being insufficient binding
at this order to produce diquark bound states.

Further insight is provided by solving the inhomogeneous Bethe-Salpeter
equation for the $0^+$ diquark vertex in the quark-condensed phase.  At
$\mu=0$ and zero total momentum: $P=0$, the additional contributions to the
quark-quark scattering kernel generate an enhancement in the magnitude of the
scalar functions in the Bethe-Salpeter amplitude.  However, as $P^2$ evolves
into the timelike region, the contributions become repulsive and block the
formation of a diquark bound state.  Conversely, increasing $\mu$ at any
given timelike-$P^2$ yields an enhancement in the magnitude of the scalar
functions, and as $\mu\to \mu_{c}^{B,\Delta=0}$ that enhancement becomes
large, which suggests the onset of an instability in the quark-condensed
vacuum.

\section{Epilogue}
We have studied a confining model of QCD using a truncation of the
Dyson-Schwinger equations that describes well the $\pi$-$\rho$ mass splitting
at $(T,\mu)=0$ and ensures that no diquark bound states appear in the
spectrum.  Employing a criterion of maximal pressure, we observe a first
order transition to a chiral symmetry breaking superfluid ground state, which
occurs at a chemical potential approximately two-thirds that required to
completely eliminate the quark condensate in the absence of diquark
condensation.  Without fine-tuning, the superfluid gap at the transition is
large, approximately one-half of that characterising quark condensation.
Thus, while completely changing the qualitative nature of the bound state
spectrum in the model; i.e., eliminating unobserved coloured bound states,
our vertex-corrected gap equation yields a phase diagram that is
semi-quantitatively the same as that obtained using the rainbow truncation.
This bolsters our confidence in the foundation of current
speculations\cite{wilczek} about the phases of high-density QCD.

The procedure we used to improve the gap equation is equally applicable to
two-colour QCD, which we analysed with the help of a pedagogical model for
the dressed-gauge boson propagator.  Diquark bound states must exist in
QC$_2$D because they are the baryons of the theory, and the truncation
procedure ensures this, with the result that flavour-nonsinglet $J^{P=\,\mp}$
mesons are degenerate with $J^{\pm}$ diquarks.  Using a straightforward,
constructive approach, we saw that at $\mu=0$ there are five Goldstone modes
in QC$_2$D, and that one of them survives at $\mu\neq 0$.  A nonzero
current-quark mass opposes diquark condensation but for light-fermions there
is always a value of the chemical potential at which a transition to the
superfluid phase takes place.  Our model studies indicate that in some
respects; such as the transition point and magnitude of the gap, the phase
diagram of QC$_2$D is quantitatively similar to that of QCD.  This
observation can be useful because the simplest superfluid order parameter is
gauge invariant in QC$_2$D and the fermion determinant is real and positive,
which makes tractable the exploration of superfluidity in QC$_2$D using
numerical simulations of the lattice theory\cite{hands}.  The results of
those studies can then be a reliable guide to features of QCD.

\vspace*{-\baselineskip}

\section*{Acknowledgments}
We acknowledge helpful interactions with D.~Blaschke, K.~Rajagopal and B. van
den Bossche.  This work was supported by the US Department of Energy, Nuclear
Physics Division, under contract number W-31-109-ENG-38, the National Science
Foundation under grant no. INT-9603385, and benefited from the resources of
the National Energy Research Scientific Computing Center.  S.M.S. is a
F.-Lynen Fellow of the A.v. Humboldt foundation.

\vspace*{-\baselineskip}


\end{document}